\documentclass[twocolumn,pra,aps,showpacs]{revtex4}
\usepackage{amsmath}
\usepackage{color}%
\usepackage{graphicx}
\usepackage{CJK}

\begin{document}

\begin{CJK*}{GBK}{song}

\title{ Photon transport mediated by an atomic chain trapped along a photonic crystal waveguide}

\author{Guo-Zhu Song$^{1}$, Ewan Munro$^{2}$, Wei Nie$^{3}$, Leong-Chuan Kwek$^{2,4,5,6}$, Fu-Guo Deng$^{7,8}$, and Gui-Lu Long$^{1,9,10}$\footnote{Corresponding author: gllong@tsinghua.edu.cn } }

\address{$^{1}$State Key Laboratory of Low-Dimensional Quantum Physics and Department of Physics, Tsinghua University, Beijing 100084, China\\
$^{2}$Centre for Quantum Technologies, National University of Singapore, 3 Science Drive 2, Singapore 117543\\
$^{3}$Institute of Microelectronics, Tsinghua University, Beijing 100084, China\\
$^{4}$Institute of Advanced Studies, Nanyang Technological University, Singapore 639673\\
$^{5}$National Institute of Education, Nanyang Technological University, Singapore 637616\\
$^{6}$MajuLab, CNRS-UNS-NUS-NTU International Joint Research Unit, UMI 3654, Singapore\\
$^{7}$Department of Physics, Applied Optics Beijing Area Major Laboratory, Beijing Normal University, Beijing 100875, China\\
$^{8}$NAAM-Research Group, Department of Mathematics, Faculty of Science, King Abdulaziz University,  Jeddah 21589, Saudi Arabia\\
$^{9}$Tsinghua National Laboratory of Information Science and Technology, Beijing 100084, China\\
$^{10}$Collaborative Innovation Center of Quantum Matter, Beijing 100084, China
}
\date{\today }

\begin{abstract}
We theoretically investigate the transport properties of a weak
coherent input field scattered by an ensemble of $\Lambda$-type
atoms coupled to a one-dimensional photonic crystal waveguide. In
our model, the atoms are randomly located in the lattice along the
crystal axis. We analyze the transmission spectrum mediated by the
tunable long-range atomic interactions, and observe the highest-energy dip.
The results show that the highest-energy dip location is associated with
the number of the atoms, which provides an accurate measuring tool for the
emitter-waveguide system. We also quantify the influence of a Gaussian
inhomogeneous broadening and the dephasing on the transmission spectrum,
concluding that the highest-energy dip is immune to both the inhomogeneous
broadening and the dephasing. Furthermore, we study photon-photon
correlations of the reflected field and observe quantum beats.
With tremendous progress in coupling atoms to photonic crystal waveguides,
our results may be experimentally realizable in the near future.
\end{abstract}

\maketitle

\section{Introduction}

In the past decades, realizing strong interactions between photons and
atoms is of central importance for  quantum optics and quantum
information processing
\cite{Duan2001Na,Kimble2008na,Chang2014,AReiserer2015}. One primary
method is to couple single atoms to high-finesse optical
microcavities \cite{Mabuchi,Birnbaum2005,Walther}, i.e., cavity
quantum electrodynamics (QED). Recently, one-dimensional (1D)
waveguide provides another promising platform for photon-atom
interactions
\cite{ShenOL2005,ShenPRL2007,PLodahl2015rmp,Diby2017rmp,Garcia2017}. In
practice, a quasi-1D waveguide can be realized by a number of
different systems, such as optical nanofibers
\cite{DayanScience2008,Vetsch2010prl,Zoubi2010njp,AngelakisPRL2011,RausPRL2011,DReitz2013PRL,AngelakisPRL2013,RYalla2014prl,FLKien2014,CSayrin2015,PSolano2017,Yan2018,Qi2018,arxiv2018},
diamond waveguide
\cite{BabinecNat2010,ClaudonPhoton2010,Sipahigil2016,MKBhaskar2017}, photonic
crystal waveguide (PCW) \cite{JDJoan2008book}, surface plasmon
nanowire \cite{AkimovNature2007,Chang2007nap,Akselrod2014,Huanglight2014,Fanglight2015}, and
superconducting microwave transmission lines
\cite{WallraffNature2004,ShenPRL2005,Abdumalikov2010,AstafievScience2010,Hoi2011,Jia2011,HoiPRL2012,LooSci2013,Jia2017,GuPR2017}.
Due to nontrivial dispersion relation caused by the periodic
dielectric structure, PCWs have attracted much attention. In the
past decade, a great progress has been made to interface atoms or
solid-state emitters with PCWs
\cite{ShenPRA2007,TLHansen2008,ALaucht2012,JDThompson2013,MArcari2014prl,TGTiecke2014,AGobannatc2014,SPYuapl2014,AGoban2015PRL,AJavadi2015natc,Sollner2015,JDHood2016PNAS,JSDouglas2015,TudelaNAT2015,CLHung2013njp,jsDoug2016prx,EWAN,Manzoni2017Natc,YLiu2017,Litao2018}.

PCWs are periodic dielectric structures in which the field
propagation can be drastically modified due to the photonic band
gaps \cite{JDJoan2008book}. Recently, the atom-light interactions in
PCWs have been explored, and rich phenomena are predicted to emerge.
Particularly, when the transition frequency of an atom lies in a
band gap, it can no longer emit a propagating photon into the
dielectric structure. However, an evanescent field forms around the
atomic position, which may be shown to exhibit the properties of an
effective cavity mode. In turn, this cavity mode can mediate effective
dipole-dipole interactions between different atoms, with a range and
strength that may be tuned experimentally \cite{SJohn1990prl,Kofman1994jmo,JSDouglas2015,jsDoug2016prx,Albrecht2017njp}.
The combined atom-PCW system then represents a novel platform for the study of
quantum many-body physics and non-linear optics \cite{JDHood2016PNAS,EWAN,Manzoni2017Natc}.


Inspired by recent developments in coupling atoms to PCWs in
experiment \cite{JDThompson2013,AGobannatc2014,SPYuapl2014,TGTiecke2014,AGoban2015PRL,JDHood2016PNAS},
we specifically study the dynamics of a weak coherent field
propagating through a $\Lambda$-type atomic ensemble coupled to a 1D
PCW. Since the precise manipulation of the atomic positions is
challenging in interfacing atoms with PCWs in experiment, we
consider the case that atoms are randomly placed in the lattice
sites along the PCW. Here, we adopt the average values from a large
sample of atomic spatial distributions and study the transport
properties of the emitter-waveguide system.

In this work, we first study the transmission properties of a weak
coherent input field and observe the highest-energy dip, which is
different from the similar case in a conventional waveguide
\cite{SongPRA}. The results reveal that the frequency of the
highest-energy dip is related to the number of the
atoms, which offers an experimental characterisation tool for the emitter-PCW
system. We also analyze the influence of the inhomogeneous broadening
in the atomic resonant transition, and quantify the effect of the dephasing in the
two lower energy levels. We conclude that the highest-energy dip is immune to both the
inhomogeneous broadening and the dephasing. Besides, since the number
of the atoms located in the lattice sites may be not fixed in experiment,
we study the transmission spectrum of the input field when the number of the
atoms is drawn from a Poisson distribution. Under this condition,
when the interaction length is much larger than the lattice constant,
some almost equally spaced dips appear in the region around the
maximum resonance frequency in the transmission spectrum,
via which we can infer the strength of the long-range atomic
interaction. While, when the interaction length is of the order
of the lattice constant, a broad dip appears in the transmission
spectrum, and the maximum resonance frequency scales
linearly with the mean number of the atoms. That is, even
though the number of the atoms follows a Poisson distribution,
we can also infer the mean number of the atoms from the maximum
resonance frequency in the transmission spectrum. Finally,
we calculate the photon correlation function of the reflected field
at its maximum resonance frequency and observe strong initial bunching.
Moreover, quantum beats emerge in the photon-photon
correlation function of the reflected field, which arises from the
long-range atomic interactions.

The paper is organized as follows. In Sec. \ref{MODEL}, we briefly
review the physics of atoms coupled to a 1D PCW, and introduce an
effective Hamiltonian for our system that an array of $\Lambda$-type
atoms is coupled to a PCW. In Sec. \ref{RESULTS}, we calculate the
transmission properties of a weak coherent input field and analyze the
influence of the inhomogeneous broadening and the dephasing.  We also
discuss the transport properties when the number of the atoms follows
a Poisson distribution, and compute the
photon-photon correlations of the reflected field. Finally, we
summarize the main results and emphasize the advantage of our system
in Sec. \ref{discussion}.

\begin{figure}
\centering\includegraphics[width=7.5cm]{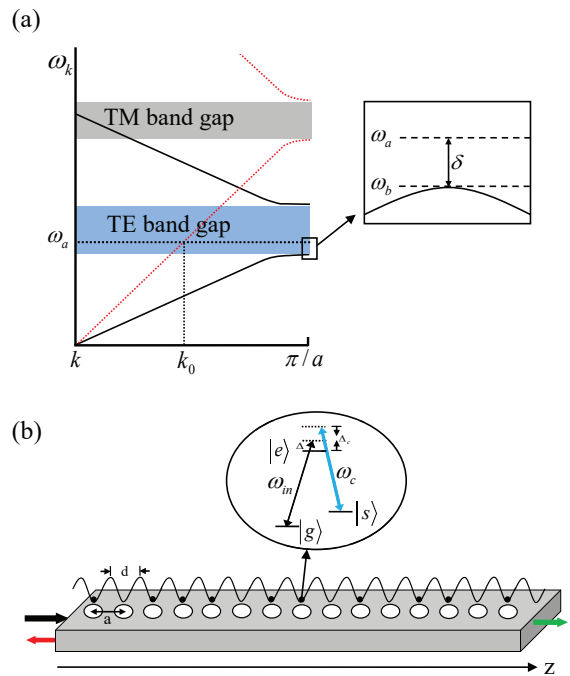} \caption{(a) Schematic
band structure of the TE and TM modes in a PCW, illustrating the
guided mode frequency $\omega_{k}$ versus the Bloch wavevector $k$.
The atomic resonance frequency $\omega_{a}$ (horizontal black dashed
line) lies in the band gap (blue region) of the TE mode (black solid
line), and is close to the lower band edge frequency $\omega_{_{b}}$
with detuning $\delta=\omega_{a}-\omega_{_{b}}$. (b) Schematic diagram
for the transport of an incident field through an atomic chain
(black dots) coupled to a PCW with  unit cell length $a$. A coherent
field (black arrow) is incident from left to scatter with the atomic
chain, which produces a reflected field (red arrow) and a
transmitted field (green arrow). The wavy line represents the
optical lattice with periodicity $d$. Due to atomic collisions
during the loading process \cite{Schlosser2002}, there exists either
no atom or only one atom in each trap site \cite{HLsorensen2016}. }
\label{figure1}
\end{figure}

\section{MODEL AND HAMILTONIAN}  \label{MODEL}

A PCW is a periodic dielectric material with regularly alternating
refractive index  \cite{JDJoan2008book,FengSCM2012,GanL2015}. For some frequencies, the
light incident into the dielectric is reflected, and the PCW acts
like a mirror. This results in the presence of band gaps in the
dispersion relation, as shown in Fig. \ref{figure1}(a). The photonic
structure supports multiple modes, e.g., the transverse-electric
(TE) and transverse-magnetic (TM) modes. Here, we consider an array
of $\Lambda$-type atoms randomly trapped in an optical lattice along
the PCW with unit cell length $a$, as shown in Fig.
\ref{figure1}(b). The atom has three relevant electronic levels,
i.e., the ground state $|\text{g}\rangle$, the metastable state $|s\rangle$
and the excited state $|e\rangle$. We assume that the resonance
frequency $\omega_{a}$ of the transition
$|\text{g}\rangle\leftrightarrow|e\rangle$ with wavevector $k_{_{0}}$ lies in
the TE band gap, and is close to the lower band edge with detuning
$\delta=\omega_{a}-\omega_{_{b}}$. In this case, due to the van Hove singularity in
the density of states, the atomic transition
$|\text{g}\rangle\leftrightarrow|e\rangle$ is dominantly coupled to the PCW modes close to
the lower band edge. In the effective mass approximation, the dispersion relation is quadratic $\omega_{k}\approx\omega_{b}(1-\alpha(k-k_{b})^{2}/k_{b}^{2})$, where
$k_{b}=\pi/a$ is the wavevector at the band edge of the TE mode and $\alpha$ characterizes the band curvature \cite{SJohn1990prl}. Since the detuning to any other band edge is
assumed to be much larger than $\delta$, we can ignore their
influence. When such an atom coupled to the PCW modes is excited at
a frequency in the band gap, it will not radiate a propagating
photon into the dielectric structure but seeds an
exponentially decaying localized photonic cloud around the atom. It
has been demonstrated that this photonic cloud has the same
properties as a real cavity mode \cite{JSDouglas2015}, which
mediates the excitations exchange with other atoms via virtual
photons. Since the band gaps of different modes occurs at different
frequencies and $\omega_{a}$ is situated far from TM band gap, we
may use the TM mode to probe the above long-range atomic
interactions. We assume that the atomic transition
$|\text{g}\rangle\leftrightarrow|e\rangle$ can couple with the TM mode via
evanescent fields, and the single-atom coupling strength is denoted
by $\Gamma_{_{1D}}$. In addition, the transition
$|e\rangle\leftrightarrow|s\rangle$ is driven by a classical control
field with the Rabi frequency $\Omega_{c}$. The system composed of
the atomic chain and the PCW can be described by an effective
non-Hermitian Hamiltonian
\cite{Chang2012,JSDouglas2015,JDHood2016PNAS,EWAN}
\begin{eqnarray}     \label{eqa1}       
\begin{split}
H_{non}\!\!=\!&-{{\sum\limits_{j}^n}}\big[(\Delta\!+\!i\Gamma_{e}^{'}/2)\sigma_{ee}^{j}\!+\!(\Delta-\Delta_{c}) \sigma_{ss}^{j}\\
&+\!\Omega_{c}(\sigma_{es}^{j}+H.c.)\big]\!-\!i\frac{\Gamma_{_{1D}}}{2}{{\sum\limits_{j,k}^n}}e^{ik_{_{0}}|z_{_{j}}-z_{_{k}}|}\sigma_{e\text{g}}^{j}\sigma_{\text{g}e}^{k}\\
&+\mathcal {J}{{\sum\limits_{j,k}^n}}\cos(k_{_{b}} z_{_{j}})\cos(k_{_{b}} z_{_{k}})e^{-|z_{_{j}}-z_{_{k}}|/L}\sigma_{e\text{g}}^{j}\sigma_{\text{g}e}^{k},
\end{split}
\end{eqnarray}
where $\Delta\!=\!\omega_{_{in}}-\omega_{a}$ is the detuning between the
frequency $\omega_{_{in}}$ of the incident field with wavevector $k_{_{in}}$
and the atomic resonance frequency $\omega_{a}$. $\Gamma_{e}^{'}$ represents
the decay rate of the state $|e\rangle$ into free space, and $z_{_{j}}$ is the position of the $j$th
atom. $\Delta_{c}\!=\!\omega_{c}-\omega_{es}$ is the detuning between the
frequency $\omega_{c}$ of the classical control field and the frequency $\omega_{es}$
of the atomic transition $|e\rangle\!\leftrightarrow\!|s\rangle$. $\mathcal {J}$ and $L$
denote the strength and characteristic length of the long-range interaction, respectively.
We can tune the strength $\mathcal {J}$ and characteristic length $L$ by adjusting the
band curvature near the lower band edge and the frequency detuning $\delta$ between the
atomic transition and the band edge \cite{JSDouglas2015}. In the last term of Eq. (\ref{eqa1}),
we consider the self-interaction part ($j\!=\!k$), which can be compensated by an external potential in experiment \cite{Manzoni2017Natc}.

Here, we study the transport of a continuous weak coherent incident
field propagating through the atomic chain. The corresponding
driving is described by $H_{dri}\!=\!\sqrt{\frac{c\Gamma_{_{1D}}}{2}}\mathcal
{E}{{\sum\limits_{j}^n}}(\sigma_{e\text{g}}^{j}e^{ik_{in}z_{_{j}}}+\sigma_{\text{g}e}^{j}e^{-ik_{in}z_{_{j}}})$,
where $\mathcal {E}$ is the
amplitude of the weak input field \cite{CanevaNJP2015}. Thus, the
properties of our system are govern by the total Hamiltonian
$H\!=\!H_{non}+H_{dri}$, and the initial state is prepared in the
global atomic ground state $|\psi_{_{0}}\rangle=|\text{g}\rangle^{\otimes n}$.
When the input field is sufficiently weak, i.e.,
$\sqrt{\frac{c\Gamma_{_{1D}}}{2}}\mathcal {E}\!\ll\!\Gamma_{e}^{'}$,
quantum jumps can be neglected \cite{EWAN}. Once the atomic dynamics
are governed by the evolution under the Hamiltonian $H$, the
transmitted ($T$) and reflected ($R$) fields can be recovered using
the input-output relations \cite{CanevaNJP2015}
\begin{eqnarray}     \label{eqa2}       
\begin{split}
a_{_{out,T}}(z) =&\; \mathcal {E}e^{ik_{in}z}+i\sqrt{\frac{\Gamma_{_{1D}}}{2c}}{{\sum\limits_{j}^n}}\sigma_{\text{g}e}^{j}e^{ik_{_{0}}(z-z_{_{j}})},\\
a_{_{out,R}}(z) =&\; i\sqrt{\frac{\Gamma_{_{1D}}}{2c}}{{\sum\limits_{j}^n}}\sigma_{\text{g}e}^{j}e^{-ik_{_{0}}(z-z_{_{j}})}.
\end{split}
\end{eqnarray}
Therefore, the transmission of the weak incident field for the steady state is given by
\begin{eqnarray}     \label{eqa3}       
\begin{split}
T=\!\frac{\langle\psi|a_{_{out,T}}^{\dagger}a_{_{out,T}}|\psi\rangle}{\mathcal {E}^{2}},
\end{split}
\end{eqnarray}
where $|\psi\rangle$ is the steady-state wavevector. For the reflected field, the equation is similar.

\section{RESULTS} \label{RESULTS}

\begin{figure}
\centering\includegraphics[width = 8.3cm]{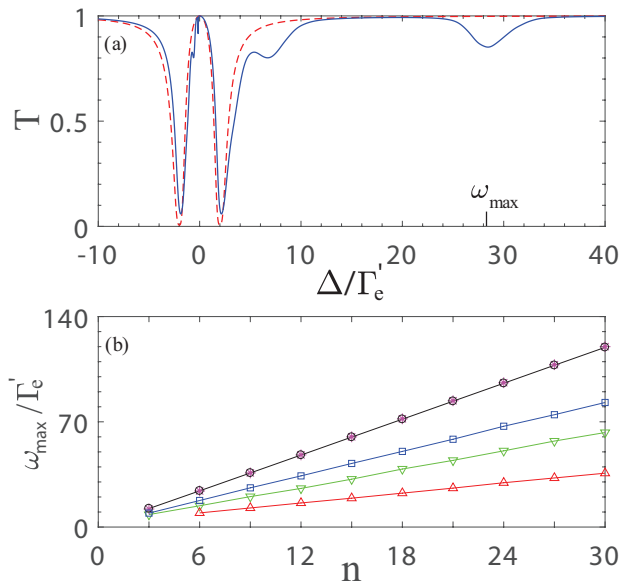} \caption{(a) The
transmission spectra of the input field for $\mathcal {J}=0$ (red
dotted line) and $\mathcal {J}=4\Gamma_{e}'$ (blue solid line). For
each case, $n=10$ atoms are randomly placed in a lattice of $N=200$
sites, and we adopt the characteristic length of the long-range
interaction as $L=100d$. (b) The maximum
resonance frequency $\omega_{max}$ versus the number $n$ of atoms
randomly placed in $N\!=\!200$ sites for $L=10^{4}d$ (black
circles), $L\!=\!100d$ (blue squares), $L\!=\!50d$ (green down triangles) and $L\!=\!20d$ (red up triangles) with $\mathcal {J}=4\Gamma_{e}'$.
The purple asterisks denote the values of $n\mathcal {J}$. Solid line is an
interpolated fit. (a)-(b)  We average over 1000 samples of atomic
spatial distributions with the parameters
$\Gamma_{_{1D}}\!=\!0.3\Gamma_{e}'$, $\mathcal
{E}=0.0001\sqrt{\frac{\Gamma_{_{1D}}}{2c}}$, $\Delta_{c}\!=\!0$, $k_{_{0}}d\!=\!\pi/2$,
$a=d$, and $\Omega_{c}\!=\!2\Gamma_{e}'$. } \label{figure2}
\end{figure}

\subsection{The transmission properties of the coherent input field} \label{TR}

Here, we study the transmission spectrum of the weak input field for $n=10$
three-level atoms randomly located in a lattice of $N=200$ sites
along a PCW, as shown in Fig. \ref{figure2}. In our simulations, we
set the lattice constant $d$ to satisfy $k_{_{0}}d\!=\!\pi/2$,
which minimizes reflection from the array
\cite{Chang2011njp,CanevaNJP2015,jsDoug2016prx,EWAN,Albrecht2017njp}.
Assuming that the input field is monochromatic, we consider two
cases: one is for $\mathcal{J}\!=\!0$, i.e., a conventional waveguide, and the other is for
$\mathcal {J}\!=\!4\Gamma_{e}'$, i.e., a PCW. We find that, for a
conventional waveguide ($\mathcal {J}\!=\!0$), electromagnetically
induced transparency (EIT) phenomenon can be observed in the
transmission spectrum, as shown in Fig. \ref{figure2}(a). This is the result of the complete
destructive interference between two atomic transitions \cite{FleischhauerREV2005}. While, for a PCW ($\mathcal
{J}\!\neq\!0$), some new dips appear in the transmission spectrum due to the
long-range interaction between atoms, which correspond to resonance frequencies of
the system. Particularly, in the limit $L/d\rightarrow\infty$, e.g., $L=10^{4}d$, we
observe that the maximum resonance frequency is independent of the
atomic spatial distributions. To interpret this, we can diagonalize the
long-range atomic interaction term in the single excitation
manifold. In fact, the $n\times n$ matrix has $(n-1)$ degenerate
resonance energies 0, and one largest resonance energy
$n\mathcal {J}$. That is, we can use $n\mathcal {J}$ to
evaluate the maximum resonance frequency $\omega_{max}$
approximatively in the limit $L/d\rightarrow\infty$, i.e.,
$\omega_{max}\approx n\mathcal {J}$. As shown in Fig.
\ref{figure2}(b), we give the values of $n\mathcal {J}$ and the maximum resonance frequency $\omega_{max}$  under  the
condition $L=10^{4}d$. We observe that, the approximate
estimation $\omega_{max}\approx n\mathcal {J}$ is valid. While
for $1\!\ll \!L/d\!\ll\!\infty$, since the maximum resonance frequency
changes with the atomic spatial distribution, we adopt the
average value from a large sample of atomic spatial distributions.
As shown in Fig. \ref{figure2}(b), we give the
relation between the maximum resonance frequency $\omega_{max}$ and
the number $n$ of atoms  for  four  cases, i.e., $L=10^{4}d$,
$L=100d$, $L=50d$ and $L=20d$. We observe that, the maximum
resonance frequency $\omega_{max}$ scales linearly with the number
of the atoms in the three cases. Differently, for the same
number of the atoms, the maximum resonance frequency
$\omega_{max}$ increases with the characteristic length $L$ of the
long-range atomic interaction, which can be interpreted by
diagonalizing the long-range atomic interaction term.
Thus,  with the fixed parameters $L$, $N$ and $\mathcal {J}$, we can
infer the number $n$ of the atoms coupled to the PCW from the
maximum resonance frequency $\omega_{max}$ in the transmission spectrum.
Note that, the conclusions mentioned above still hold if the atomic resonance
frequency is close to the upper band edge, but with opposite sign of the last term in Eq. (\ref{eqa1}). This provides an effective approach to change the sign of the long-range interactions by tuning the
atomic resonance frequency close to either the upper or lower band edges.
Since the highest-energy dip
location in the transmission spectrum is related with the number of the atoms,
we provide an accurate measuring tool for the emitter-waveguide system.
In the following discussions, we will mainly focus on the condition $L=100d$,
which may be accessible in the `alligator' PCW with state-of-the-art
fabrication \cite{SPYuapl2014,JDHood2016PNAS,JSDouglas2015}. However, our conclusions will be independent of specific choice of the characteristic length in the limit $L/d\gg1$. While, for short-range interaction, e.g., $L/d\approx1$, the coupling between distant atoms is very weak and the band gap interaction is negligible.

\begin{figure}[tpb]    
\centering\includegraphics[width= 9.2cm]{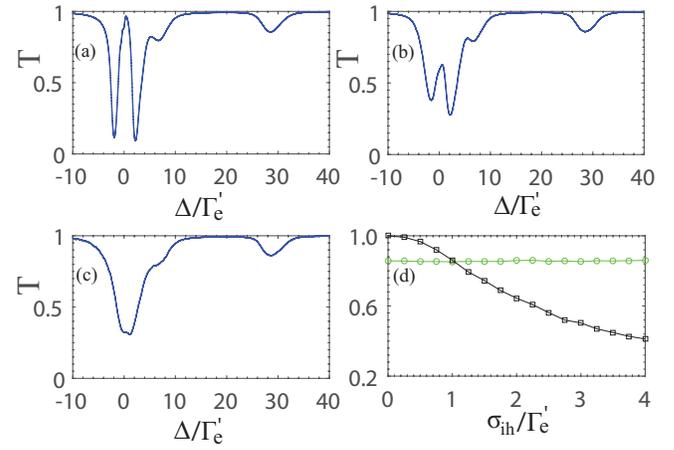} \caption{ The
transmission spectrum of the input field as a function of the
frequency detuning $\Delta/\Gamma_{e}'$ for (a)
$\sigma_{_{ih}}\!=\!0.5\Gamma_{e}'$, (b)
$\sigma_{_{ih}}\!=\!2.0\Gamma_{e}'$, (c)
$\sigma_{_{ih}}\!=\!5.4\Gamma_{e}'$. (d) The transmission $T_{peak}$
(black squares) at the EIT-like peak and the transmission $T_{dip}$
(green circles) at the maximum resonance frequency versus the
parameter $\sigma_{_{ih}}$ in the inhomogeneous broadening. Solid line
is an interpolated fit. (a)-(d) $n=10$ atoms are randomly placed in
a lattice of $N=200$ sites, and we average over $30000$ single-shot realizations with $\Gamma_{_{1D}}\!=\!0.3\Gamma_{e}'$, $\mathcal
{E}=0.0001\sqrt{\frac{\Gamma_{_{1D}}}{2c}}$, $k_{_{0}}d\!=\!\pi/2$,
$a=d$, $\Delta_{c}\!=\!0$, $L\!=\!100d$, $\mathcal
{J}=4\Gamma_{e}'$, and $\Omega_{c}\!=\!2\Gamma_{e}'$. }
\label{figure3}
\end{figure}



In above simulations, we assume that the
$\Lambda$-type atoms located in the lattice are identical. However,
in experiment, due to the off-resonant trapping fields for the atomic ensemble,
the emitters trapped in different sites suffer different vector light shifts \cite{DReitz2013PRL,Lacroute2012}. This
effect may cause the inhomogeneous broadening in the atomic transition
of the emitters. For simplicity, here we just consider the influence of the broadening in the $|e\rangle\leftrightarrow|s\rangle$ transition on the transmission spectrum.
For concreteness, we consider that the inhomogeneous broadening is Gaussian with the
probability density $\rho_{_{ih}}(\Delta_{ih})=\frac{1}{\sigma_{_{ih}}\sqrt{2\pi}}\exp({-\frac{\Delta_{_{ih}}^{2}}{2\sigma_{_{ih}}^{2}}})$,
where $\Delta_{ih}$ is the detuning from the expected frequency of the atomic transition $|e\rangle\leftrightarrow|s\rangle$
and $\sigma_{_{ih}}$ being the standard deviation is a measure of the width of the inhomogeneous broadening.
In each single-shot realization, we generate some random variables from the Gaussian probability density as
the detuning of the atoms and a random set as the positions of the atoms, via which we calculate the transmission
spectrum. As shown in Figs. \ref{figure3}(a)-\ref{figure3}(c), we give the transmission spectrum of the
input field in three cases, i.e., $\sigma_{_{ih}}\!=\!0.5\Gamma_{e}', 2.0\Gamma_{e}',
5.4\Gamma_{e}'$. The results show that the maximum resonance frequency $\omega_{max}$ does not
vary with the parameter $\sigma_{_{ih}}$. We observe that, with the increment of the
parameter $\sigma_{_{ih}}$, the transmission $T_{peak}$ at the EIT-like peak decreases,
which indicates that the inhomogeneous broadening destroys the destructive interference between the two atomic transitions
$|\text{g}\rangle\!\leftrightarrow\!|e\rangle$ and $|e\rangle\!\leftrightarrow\!|s\rangle$.
Interestingly, the transmission $T_{dip}$ at the maximum resonance frequency almost does
not change with the parameter $\sigma_{_{ih}}$.  Moreover, we find that, when the parameter
$\sigma_{_{ih}}$ is sufficiently large, e.g., $\sigma_{_{ih}}\!=\!5.4\Gamma_{e}'$, the EIT-like
peak will almost completely disappear, as shown in Fig. \ref{figure3}(c).
For clear presentation, we give the transmission $T_{peak}$ at the
EIT-like peak and the transmission $T_{dip}$ at the maximum resonance frequency as a
function of the parameter $\sigma_{_{ih}}$. As shown in Fig. \ref{figure3}(d), we
find that the EIT-like peak are sensitive to the parameter $\sigma_{_{ih}}$.
While, the existence of the highest-energy dip is immune to the inhomogeneous broadening.
That is, the property originating from the long-range atomic interaction term in Eq. (\ref{eqa1})
is not influenced by the inhomogeneous broadening of the transition $|e\rangle\leftrightarrow|s\rangle$.
In other words, even though the parameter $\sigma_{_{ih}}$ in the
inhomogeneous broadening is much larger than the decay rate $\Gamma_{e}'$,
we can also observe the highest-energy dip clearly in the transmission
spectrum and acquire the maximum resonance frequency $\omega_{max}$.

\begin{figure}[tpb]    
\centering\includegraphics[width=8.8cm]{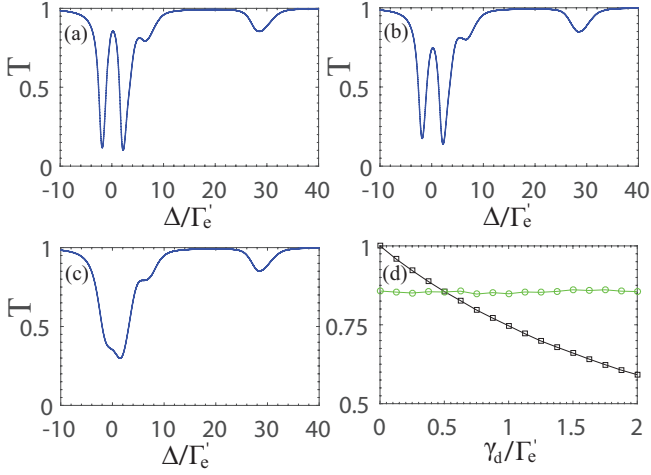}
\caption{ The transmission spectrum of the coherent incident field as a function of the frequency detuning $\Delta/\Gamma_{e}'$ for (a) $\gamma_{_{d}}\!=\!0.5\Gamma_{e}'$, (b) $\gamma_{_{d}}\!=\!1.0\Gamma_{e}'$, (c) $\gamma_{_{d}}\!=\!5.5\Gamma_{e}'$. (d) The transmission $T_{peak}$ (black squares) at the EIT-like peak and the transmission $T_{dip}$ (green circles) at the maximum resonance frequency versus the dephasing rate $\gamma_{_{d}}$. Solid line is an interpolated fit. (a)-(d) $n\!=\!10$ atoms are randomly located in a lattice of $N=200$ sites, and we average over 1000 samples of atomic spatial distributions with $\Gamma_{_{1D}}\!=\!0.3\Gamma_{e}'$, $\mathcal {E}=0.0001\sqrt{\frac{\Gamma_{_{1D}}}{2c}}$, $k_{_{0}}d\!=\!\pi/2$, $a=d$, $\Delta_{c}\!=\!0$, $L\!=\!100d$, $\mathcal {J}=4\Gamma_{e}'$, $\sigma_{_{ih}}\!=\!0$, and $\Omega_{c}\!=\!2\Gamma_{e}'$. } \label{figure4}
\end{figure}

\begin{figure} [tpb]    
\centering\includegraphics[width=9.1cm]{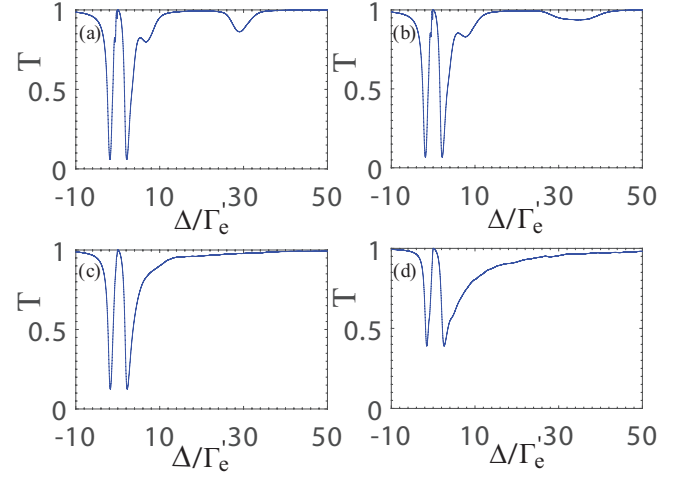}
\caption{ (a) The transmission spectrum of the incident field as a function of the detuning $\Delta/\Gamma_{e}'$ for different decay rates (a) $\gamma_{em}=0.0001\Gamma_{e}'$, (b) $\gamma_{em}=0.001\Gamma_{e}'$, (c) $\gamma_{em}=0.01\Gamma_{e}'$, (d) $\gamma_{em}=0.1\Gamma_{e}'$.
(a)-(d) $n\!=\!10$ atoms are randomly located in a lattice of $N=200$ sites, and we average 1000 samples of atomic positions with the parameters $\Gamma_{_{1D}}\!=\!0.3\Gamma_{e}'$, $\mathcal {E}=0.0001\sqrt{\frac{\Gamma_{_{1D}}}{2c}}$, $k_{_{0}}d\!=\!\pi/2$, $L\!=\!100d$, $a=d$, $\Delta_{c}\!=\!0$, $\sigma_{_{ih}}\!=\!0$, $\gamma_{_{d}}\!=\!0$, $\mathcal {J}=4.0\Gamma_{e}'$, and $\Omega_{c}\!=\!2\Gamma_{e}'$. } \label{figure5}
\end{figure}

\begin{figure*} [tpb]    
\begin{center}
\centering\includegraphics[width= 16.5cm]{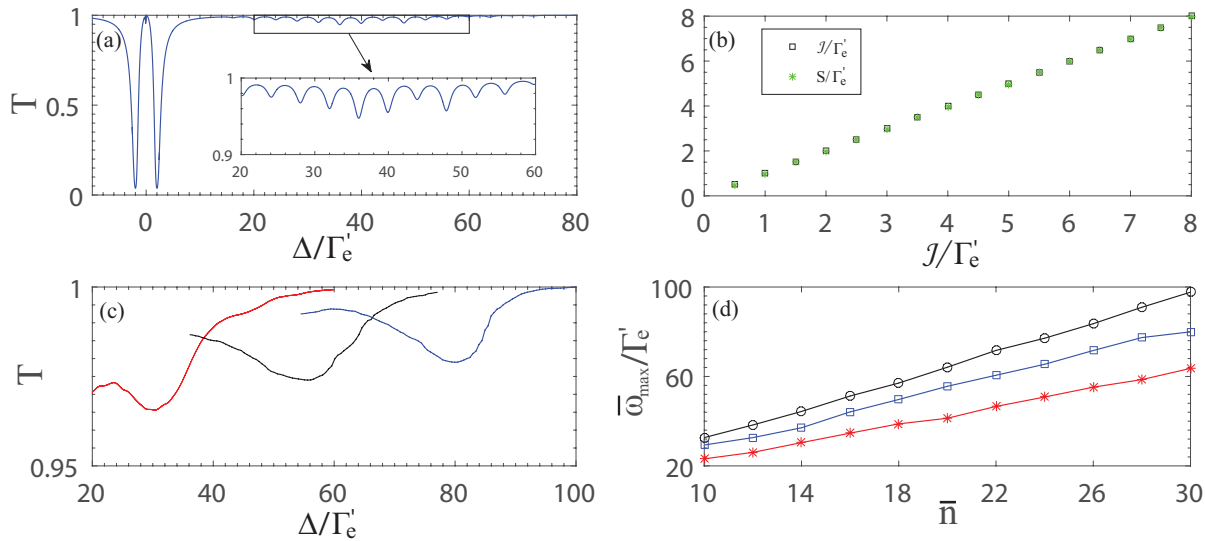}
\caption{(a) The transmission spectrum of the input field as a function of the frequency detuning $\Delta/\Gamma_{e}'$.  (b) The mean spacing $S$ between the adjacent dips as a function of the long-range interaction strength $\mathcal {J}$. (c) The transmission spectra of the input field in the region around the maximum resonance frequency for \={n}=10 (red line), 20 (black line), 30 (blue line) with $L\!=\!100d$. (d) The maximum resonance frequency $\overline{\omega}_{max}$ versus the mean number \={n} of atoms for $L\!=\!200d$ (black circles), $L\!=\!100d$ (blue squares) and $L\!=\!50d$ (red asterisks).
(a)-(b) The number of the atoms is drawn from a Poisson distribution of mean \={n}=10 with $L=10^{4}d$; (a) and (c)-(d) $\mathcal {J}=4\Gamma_{e}'$; (a)-(d) we average over 30000 single-shot samples with the parameters
$\Gamma_{_{1D}}\!=\!0.3\Gamma_{e}'$, $\mathcal {E}=0.0001\sqrt{\frac{\Gamma_{_{1D}}}{2c}}$, $N=200$, $k_{_{0}}d\!=\!\pi/2$, $a=d$, $\Delta_{c}\!=\!0$, $\sigma_{_{ih}}\!=\!0$, $\gamma_{_{d}}\!=\!0$, $\gamma_{em}\!=\!0$, and $\Omega_{c}\!=\!2\Gamma_{e}'$.
 }   \label{figure6}
\end{center}
\end{figure*}

In practical emitter-waveguide systems, the transport
properties of the coherent input field may be influenced by the dephasing of the two lower energy levels.
The dephasing rates of the atoms coupled to a PCW have not been reported yet. While, in a similar system, i.e., atoms trapped in the surface of the optical nanofibers, the dephasing of the two energy levels $|\text{g}\rangle$ and $|s\rangle$ exists due to temperature dependent light
shifts in the optical trap \cite{DReitz2013PRL}, thermal motion of the atoms \cite{CSayrin2015,HLsorensen2016} and atom-dependent Larmor precession caused by the residual magnetic field \cite{GoutaudPRL2015}. For simplicity, we assume
that the dephasing rates of all atoms are identical. The dynamics
of the atomic system is governed by the master equation for the
atomic density operator:
\begin{eqnarray}     \label{eqa4}       
\frac{d\rho}{dt}=\!\!\!\!\!&&-i[H_{I},\rho]-\frac{\Gamma_{e}^{'}}{2}{{\sum\limits_{j}^n}}\big(\{\sigma_{ee}^{j},\rho\}-2\sigma_{\text{g}e}^{j}\rho \sigma_{e\text{g}}^{j}\big)\nonumber\\
&&-\frac{\gamma_{_{d}}}{2}{{\sum\limits_{j}^n}}\big(\{\sigma_{ss}^{j},\rho\}\!-\!2\sigma_{ss}^{j}\rho \sigma_{ss}^{j}\!+\!\{\sigma_{\text{g}\text{g}}^{j},\rho\}\!-\!2\sigma_{\text{g}\text{g}}^{j}\rho \sigma_{\text{g}\text{g}}^{j}\big)\nonumber\\
&&-\frac{\Gamma_{_{1D}}}{2}{{\sum\limits_{j,k}^n}}\cos({k_{_{0}}|z_{_{j}}\!-\!z_{_{k}}|})(\sigma_{e\text{g}}^{j}\sigma_{\text{g}e}^{k}\rho+\rho \sigma_{e\text{g}}^{j}\sigma_{\text{g}e}^{k}\nonumber\\
&&-2\sigma_{\text{g}e}^{k}\rho \sigma_{e\text{g}}^{j}),
\end{eqnarray}
where
\begin{eqnarray}     \label{eqa5}       
H_{I}=\!\!\!\!\!&&-{{\sum\limits_{j}^n}}\big[\Delta \sigma_{ee}^{j}+(\Delta-\Delta_{c}) \sigma_{ss}^{j}+\Omega_{c}(\sigma_{es}^{j}+H.c.)\big]\nonumber\\
&&+\mathcal {J}{{\sum\limits_{j,k}^n}}\cos(k_{_{b}} z_{_{j}})\cos(k_{_{b}}z_{_{k}})e^{-|z_{_{j}}-z_{_{k}}|/L}\sigma_{e\text{g}}^{j}\sigma_{\text{g}e}^{k}\nonumber\\
&&+\frac{\Gamma_{_{1D}}}{2}{{\sum\limits_{j,k}^n}}\sin({k_{_{0}}|z_{_{j}}-z_{_{k}}|})\sigma_{e\text{g}}^{j}\sigma_{\text{g}e}^{k}.
\end{eqnarray}
In Eq. (\ref{eqa4}), the third term on the right-hand
side describes the dephasing of
the two lower states $|\text{g}\rangle$ and $|s\rangle$. We assume that the dephasing rates of the two levels are the same and are given by $\gamma_{_{d}}$. As shown in Fig.
\ref{figure4}, we calculate the effect of the dephasing on the transmission spectrum. Here, we consider three
choices of the dephasing rate, i.e.,
$\gamma_{_{d}}=0.5\Gamma_{e}',1.0\Gamma_{e}',5.5\Gamma_{e}'$. The results show that the maximum resonance frequency $\omega_{max}$ does not
change with $\gamma_{_{d}}$. While, when the dephasing rate
is changed from $0.5\Gamma_{e}'$ to $1.0\Gamma_{e}'$,
the transmission at the EIT-like peak decreases and the highest-energy dip remains unchanged, as shown in
Figs. \ref{figure4}(a)-\ref{figure4}(b).
Interestingly, when the dephasing rate is sufficiently large, e.g., $\gamma_{_{d}}=5.5\Gamma_{e}'$, the EIT-like peak will almost completely disappear. That is, similar to the influence of inhomogeneous broadening, the dephasing of the two lower energy levels can also destroy the destructive interference between the two atomic transitions.
Additionally, to show the phenomenon
more clearly, we give the transmission $T_{peak}$ at the EIT-like
peak and the transmission $T_{dip}$ at the maximum resonance
frequency as a function of the dephasing rate $\gamma_{_{d}}$. As
shown in Fig. \ref{figure4}(d), we observe that the depth of
the highest-energy dip is immune to the dephasing rate.
That is, even though the dephasing effect in the two lower energy
levels is strong, we can observe the highest-energy dip
in the transmission spectrum. In fact, in the system of atoms trapped along nanofibers \cite{DReitz2013PRL,CSayrin2015}, the dephasing rate is in the range of $\gamma_{_{d}}\simeq2\pi [200\text{Hz} - 50 \text{kHz}]$.
For Cs atoms coupled to PCWs, the decay rate $\Gamma_{e}'=2\pi\cdot4.56$MHz has been reported \cite{JDHood2016PNAS}. That is, $\gamma_{_{d}}/\Gamma_{e}'\simeq[4.4\cdot10^{-5}-0.01]$ has been obtained. In such range, we can observe both the highest-energy dip and EIT-like peak clearly in the transmission spectrum.


In the discussion above, we have considered the dissipation via the decay of the atomic excited state
into free space. While, for a realistic PCW in experiment, there probably also exists another one dissipation channel, i.e., the decay of the TE to TM modes \cite{JDHood2016PNAS}. Here, we assume that the decay rate of the TE to TM modes is given by $\gamma_{em}$. As shown in Fig. \ref{figure5}, we give the transmission spectrum of the input field for four choices of the decay rate, i.e., $\gamma_{em}=0.0001\Gamma_{e}',0.001\Gamma_{e}', 0.01\Gamma_{e}', 0.1\Gamma_{e}'$. The results show that the existence of the highest-energy dip is sensitive to the decay of the TE to TM modes. In detail, when the decay rate $\gamma_{em}$ is changed from $0.0001\Gamma_{e}'$ to $0.01\Gamma_{e}'$, the highest-energy dip will completely disappear. That is, to acquire the maximum resonance frequency $\omega_{max}$ in the transmission spectrum, the decay of the TE to TM modes must be strongly suppressed.

In the previous sections, we assume that the number of the atoms coupled
to the PCW is fixed, and only the atomic positions are considered to
be random. However, in practical experiment, the number of the atoms
may follow a specific distribution, such as Poisson distribution. We first consider the case $L/d\rightarrow\infty$, e.g., $L=10^{4}d$, and plot the average transmission spectrum of the
input field when the number of the atoms is randomly drawn from a
Poisson distribution with mean \={n}=10, as shown in Fig. \ref{figure6}(a). In each single-shot realization, we get a random value from the Poisson
distribution as the number of the atoms, and generate a random set as the positions of the atoms, by which we compute the transmission spectrum.
The results show that, some almost
equally spaced dips appear in the region around the maximum resonance frequency
\={n}$\mathcal {J}$ for the mean number \={n} of the atoms. As
we have discussed above, in the limit $L/d\rightarrow\infty$, the maximum resonance frequency is related with the number of the atoms, i.e., $\omega_{max}\approx n\mathcal {J}$.
Thus, in Fig. \ref{figure6}(a), each dip in the region around the maximum resonance corresponds to one
random number drawn from the Poisson distribution.
In other words, the difference between the numbers of the atoms
corresponding to the adjacent dips is one, i.e., the frequency
spacing between the nearest-neighbor dips is the strength
$\mathcal {J}$ of the long-range atomic interactions. To verify this
conclusion, we calculate the mean spacing $S$ between adjacent dips
as a function of the strength $\mathcal {J}$ of the long-range
atomic interactions, as shown in Fig. \ref{figure6}(b). For
comparison, we also plot the values of the strength $\mathcal {J}$.
We see that, the mean spacing $S$ between the adjacent dips is
almost the same as the strength $\mathcal {J}$. That is, under the condition $L/d\rightarrow\infty$, one can obtain the strength $\mathcal {J}$ of the long-range atomic interactions by calculating the mean spacing between adjacent
dips in the transmission spectrum.

\begin{figure} [tpb]    
\centering\includegraphics[width=8cm]{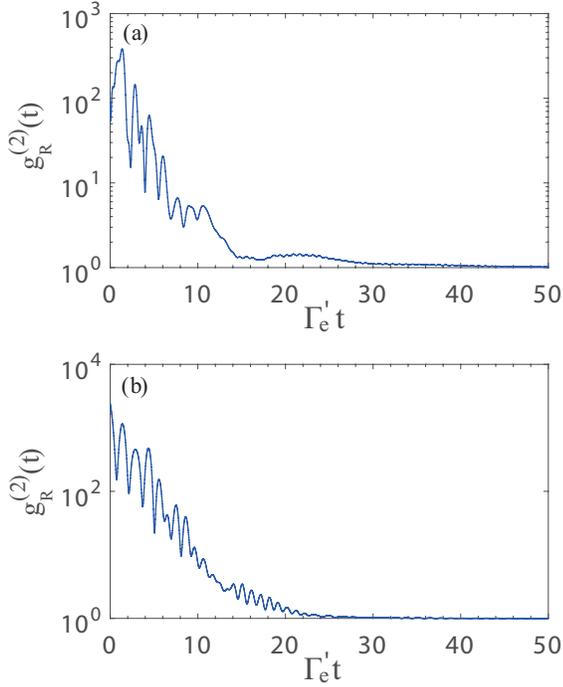}
\caption{ The photon-photon correlation function $\text{g}_{_{R}}^{(2)}$ of the
reflected field at its maximum resonance frequency $\omega_{max}$ when $n\!=\!10$ atoms are
randomly placed over $N\!=\!200$ sites with the parameters (a)
$\mathcal {J}=1.0\Gamma_{e}'$, (b) $\mathcal {J}=4.0\Gamma_{e}'$.
Here, we average 1000 samples of atomic positions with the parameters
$\Gamma_{_{1D}}\!=\!0.3\Gamma_{e}'$, $\mathcal {E}=0.0001\sqrt{\frac{\Gamma_{_{1D}}}{2c}}$, $k_{_{0}}d\!=\!\pi/2$, $L\!=\!100d$, $a=d$, $\Delta_{c}\!=\!0$, $\sigma_{_{ih}}\!=\!0$, $\gamma_{_{d}}\!=\!0$, $\gamma_{em}=0$, and $\Omega_{c}\!=\!2\Gamma_{e}'$. } \label{figure7}
\end{figure}

To proceed, we evaluate the transmission spectrum of the weak input
field under the condition $L\!=\!100d$ when the number of the atoms follows a Poisson distribution. As shown in Fig. \ref{figure6}(c), we consider three cases, i.e., the mean numbers of the atoms are \={n}=10, 20, 30, respectively. The results show that a broad dip appears in the region around the maximum resonance, which is different from the case $L/d\rightarrow\infty$. Here, we denote the highest-energy dip location as $\overline{\omega}_{max}$ when the number of the atoms follows a Poisson distribution. We observe that
$\overline{\omega}_{max}\approx \omega_{max}$ is valid in the three cases mentioned above. Moreover, we plot  $\overline{\omega}_{max}$ as a function of the mean number \={n} for three different interaction lengths, as shown in Fig. \ref{figure6}(d). We find that, the dip location $\overline{\omega}_{max}$ (i.e., the maximum resonance frequency) scales linearly with the mean number \={n} of the atoms, which is similar to the case that the number of the atoms is fixed. That is, when the number of atoms follows a Poisson distribution, we can infer \={n} from the maximum resonance frequency $\overline{\omega}_{max}$ in the transmission spectrum.

\subsection{Photon-photon correlation}  \label{correlation}

The key feature of non-classical light is the existence of correlations between photons, which can be characterized by the second-order correlation function $\text{g}^{(2)}$ (also called photon-photon correlation function) \cite{Loudon2003}. For a weak coherent state, the photon-photon correlation function $\text{g}^{(2)}$ of the output field is given by
\begin{eqnarray}     \label{eqa6}       
\text{g}_{\alpha}^{(2)}(\tau)\!\!=\!\!\frac{\langle\psi|a_{_{\alpha}}^{\dagger}(z)e^{iH\tau}a_{_{\alpha}}^{\dagger}(z)a_{_{\alpha}}(z)e^{-iH\tau}a_{_{\alpha}}(z)|\psi\rangle}{|\langle\psi|a_{_{\alpha}}^{\dagger}(z)a_{_{\alpha}}(z)|\psi\rangle|^{2}}.
\end{eqnarray}
Here, $|\psi\rangle$ denotes the steady-state wavevector, and $\alpha\!=\!T, R$.

Now, with a weak coherent incident field ($\sqrt{\frac{c\Gamma_{_{1D}}}{2}}\mathcal {E}\!\!\ll\!\!\Gamma_{e}^{'}$), we analyze the photon-photon correlations
of the output field at the corresponding maximum resonance frequency. As shown in Fig. \ref{figure7}, we compute the photon correlation function of the reflected
field with two choices of the strength $\mathcal {J}$ when $n=10$ three-level atoms are randomly located in a lattice of $N\!\!=\!\!200$ sites along a PCW. We find that,
strong initial bunching appears in the reflected field in the two cases, i.e., $\text{g}_{_{R}}^{(2)}(t\!=\!0)\gg1$. Since the reflected field originates purely
from the scattering by the atomic ensemble, the strong initial bunching $\text{g}_{_{R}}^{(2)}(t\!=\!0)\gg1$ indicates that the $\Lambda$-type atomic chain can
scatter two photons simultaneously. Comparing Fig. \ref{figure7}(a) and Fig. \ref{figure7}(b), we conclude that the correlation properties of the reflected field
is influenced by the strength $\mathcal {J}$ of the long-range atomic interactions. That is, the initial bunching becomes stronger when we enhance the long-range
interaction strength $\mathcal {J}$. Furthermore, we observe quantum beats \cite{ZhengPRL2013} in the second-order correlation function of the reflected field. Evidently, the stronger the long-range interaction, the more visible the quantum beats become. That is, the long-range atomic interactions
in our system can cause the quantum beats in the photon-photon correlation function of the reflected field. The results reveal that our emitter-PCW system may provide
an effective platform for experimental study of the nonclassical light.




\section{DISCUSSION AND SUMMARY}   \label{discussion}


In our system, the PCW is mainly characterized by the parameters $\mathcal {J}$ and $L$, which represent the characteristic strength and length of the long-range interaction, respectively. In practice, by tuning the detuning $\delta$ from the band edge and band curvature $\alpha$ at the band edge, one can control the parameters $\mathcal {J}$ and $L$. In particular, the parameters $\mathcal {J}$ and $L$ of the emitter-PCW system are given by $\mathcal {J}\!=\!\bar{g}_{c}^2/2\delta$ and
$L\!=\!\sqrt{\alpha\omega_{b}/{k_{b}^2\delta}}$, respectively \cite{JSDouglas2015}. Here $\bar{g}_{c}=g_{d}\sqrt{d/L}$, and $g_{d}$ is the vacuum Rabi splitting in a photonic crystal cavity with length $d$. Thus, by reducing the detuning $\delta$ from the band edge, we can obtain a stronger and longer-range atom-atom interaction with a fixed band curvature.

In addition to the emissions of the atomic excited state into free space at rate $\Gamma_{e}^{'}$ and the TM modes at rate $\Gamma_{_{1D}}$, photon loss at characteristic rate $\kappa$ may also exist in a realistic PCW \cite{JSDouglas2015}. This could be due to scattering and absorption loss of the photonic crystal structure. With no decay of the TE to TM modes, the total effective dissipation rate of an excited atom is $\Gamma_{tot}=\Gamma_{_{1D}}+\Gamma_{e}^{'}+\kappa(\bar{g}_{c}/2\delta)^2$. Here, the last term proportional to $\kappa$ denotes the Purcell enhancement caused by the case that an atom is off-resonantly coupled to an effective cavity mode. In order to observe the phenomena arising
from the long-rang interaction mentioned above, the interaction strength $\mathcal {J}$ must exceed the total dissipation rate $\Gamma_{tot}$. In fact, the ratio $\mathcal {J}/\Gamma_{tot}$ can be optimized by tuning the detuning $\delta$. We find that, the theoretical maximum is $\mathcal {J}/\Gamma_{tot}=\sqrt{\bar{g}_{c}^2/\kappa\Gamma}/2$ when $\delta^2=\kappa\bar{g}_{c}^2/4\Gamma$, where $\Gamma=\Gamma_{_{1D}}+\Gamma_{e}^{'}$. Note that, optimizing the detuning $\delta$ also changes the interaction length $L=\sqrt{\alpha\omega_{b}/{k_{b}^2\delta}}$. In order to keep the length $L$ fixed, we must also tune the band curvature $\alpha$.

In a recent experiment \cite{JDHood2016PNAS}, Hood \emph{et al.} experimentally
observed signatures of collective atom-light interactions by tuning the band edge
frequency of the PCW relative to cesium atoms trapped along an alligator PCW. In their experiment, at the detuning $\delta=60$GHz inside the band gap, the free space emission rate of the cesium atom is $\Gamma_{e}^{'}/2\pi\approx5.0$MHz, and the coupling strength between the TM modes and a single atom is $\Gamma_{_{1D}}/\Gamma_{e}^{'}\approx9.1\times10^{-3}$.
Specifically, with lattice constant $d=370$nm, one finds that the characteristic strength and length of the long-range interaction are $\mathcal {J}/\Gamma_{e}^{'}\approx0.182$ and $L/d\approx80$, respectively. As shown in Fig. 4 of Ref. \cite{JDHood2016PNAS}, they gave the ratio between the parameters $\mathcal {J}$ and $\Gamma_{_{1D}}$ as a function of the detuning $\delta$. While, their values of the parameters $\mathcal {J}$ and $\Gamma_{_{1D}}$ are not yet good enough to observe the results shown in this paper. In additional to the alligator PCW, another possibility is to use a slot PCW \cite{PLodahl2015rmp,para1so2015}, which is obtained by placing two plasmon waveguides next to each other. By confining
the emitters at the center of the nanostructure, one can get the values of $\mathcal {J}/\Gamma_{e}^{'}\simeq6$, $\Gamma_{_{1D}}/\Gamma_{e}^{'}\approx0.3$ for a detuning $\delta=20$GHz from the band edge, and the range of atom-atom interaction is $L/d\approx80$. In summary, to date, the emitter-PCW system has the curvature parameter $1.0\!\!\leq\!\alpha\!\leq\!\!10.6$, giving
the parameters $1.25\!\!\leq\!\mathcal {J}/\Gamma_{e}^{'}\!\leq\!\!6.0$ and $5\!\leq \!L/d\!\leq\!200$ in current experiments \cite{AGobannatc2014,SPYuapl2014,JDHood2016PNAS,JSDouglas2015}.
Thus, after optimizing the parameters of the emitter-PCW system, the values of $\mathcal {J}$ and $L$ (i.e., $\mathcal {J}/\Gamma_{e}^{'}\!=\!4$, $L/d\!=\!100$) in our calculations fall well in the experimentally achievable limits. Note that, the case where $L/d=10^{4}$ merely serves as an example for theoretical analysis of the phenomena. Nevertheless, we believe that there is a bright future
for the setups yielding great improvement on these rates.

In summary, we have theoretically studied the transport properties of a
$\Lambda$-type three-level atomic ensemble coupled to the band edge of a PCW.
Considering the precise control of the atomic positions is still
challenging in interfacing atoms with PCWs, we assume that atoms
are randomly placed in the lattice along the axis of the PCW in our
model. With the effective non-Hermitian Hamiltonian, we calculate
the transmission spectrum of a weak coherent input field and observe
the highest-energy dip, which arises from the long-range atomic
interactions. We find that, in the limit $L/d\gg1$,  the maximum
resonance frequency scales linearly with the number of the atoms
coupled to the PCW, which may provide an accurate measuring tool for
the emitter-PCW system. We also quantify the effect of a Gaussian
inhomogeneous broadening of the transition $|e\rangle\leftrightarrow|s\rangle$
and the dephasing in the two lower energy levels on the transmission spectrum.
The results reveal that the existence of the highest-energy dip is immune to both the
inhomogeneous broadening and the dephasing. Furthermore, we analyze the transmission spectrum of the weak input field when
the number of the atoms follows a Poisson distribution. We find that,  when the interaction length $L$ is of the order of the
lattice constant $d$, a broad highest-energy dip appears in the transmission spectrum. That is, even though the number of atoms follows
a Poisson distribution, we can infer the mean number \={n} of the atoms from the maximum resonance frequency.
Finally, we calculate the photon-photon correlations of the reflected field at the maximum
resonance frequency and observe strong initial bunching. Moreover, the long-range atomic interactions
cause quantum beats in the photon-photon correlation function of the reflected field.
We emphasize that, our work takes advantage of the emitter-PCW system
that one can separately tune the range and strength of the atomic
interactions by engineering the dispersion of the structure
\cite{JSDouglas2015,jsDoug2016prx}. In theory, the range of the atom-atom interaction is from effectively infinite to nearest neighbor \cite{JSDouglas2015}.
Since tremendous progress has
been made to experimentally realize the coupling between emitters
and PCWs \cite{JDThompson2013,AGobannatc2014,SPYuapl2014,TGTiecke2014,AGoban2015PRL,JDHood2016PNAS},
the results in our model may be realizable in the near future.


\section*{ACKNOWLEDGMENTS}

This work is supported by the China Postdoctoral Science Foundation under Grant No. 2017M620732, the National Natural Science Foundation of China under Grants No. 11175094 and No. 91221205, and National Basic Research Program of China under Grant No. 2015CB921002.
E.M., and L.-C.K acknowledge support from the National Research Foundation and Ministry of Education, Singapore. F.-G.D. is supported by the National Natural Science Foundation of
China under Grants No. 11474026 and No. 11674033, and the Fundamental Research Funds for the Central Universities under Grant No. 2015KJJCA01. G.-L.L. acknowledges support from the Center of Atomic and Molecular Nanosciences, Tsinghua University, and the Beijing Advanced Innovation Center for Future Chip (ICFC).

\end{CJK*}

\end{document}